\documentclass[a4paper,11pt,showkeys]{article}
\usepackage{amsmath,amssymb,amsfonts}
\usepackage{graphicx}
\usepackage{rotating}
\usepackage[hang,bf,nooneline]{subfigure}
\usepackage[hang,bf,nooneline]{caption}

\newlength{\dslashwidth}

\newcommand{\tb}{\ensuremath{\tan\beta}}

\newcommand{\bq}{\begin{equation}}
\newcommand{\eq}{\end{equation}}
\newcommand{\ba}{\begin{array}}
\newcommand{\ea}{\end{array}}
\newcommand{\bqa}{\begin{eqnarray}}
\newcommand{\eqa}{\end{eqnarray}}

\newcommand{\lnf}{{\ifmmode \Lambda^{(N_f)} \else $\Lambda^{(N_f)}$\fi}}
\newcommand{\ms}{{\ifmmode \overline{MS} \else $\overline{MS}$\fi}}
\newcommand{\dr}{{\ifmmode \overline{DR} \else $\overline{DR}$\fi}}
\newcommand{\lms}{{\ifmmode \Lambda^{(5)}_{\overline{MS}} \else $\Lambda^{(5)}_{\overline{MS}}$\fi}}
\newcommand{\lam}{{\ifmmode \Lambda \else $\Lambda$\fi}}
\newcommand{\gev}{{\ifmmode {\rm GeV} \else ${\rm GeV}$\fi}}
\newcommand{\gevc}{{\ifmmode {\rm GeV/c^2} \else ${\rm GeV/c^2}$\fi}}
\newcommand{\tev}{{\ifmmode {\rm TeV} \else ${\rm TeV}$\fi}}
\newcommand{\tevc}{{\ifmmode {\rm TeV/c^2} \else ${\rm TeV/c^2}$\fi}}
\newcommand{\lp}{{\ifmmode L^+  \else $L^+$\fi}}
\newcommand{\lm}{{\ifmmode L^-  \else $L^-$\fi}}
\newcommand{\mlp}{{\ifmmode M(L^-) \else $M(L^-)$\fi}}
\newcommand{\mlz}{{\ifmmode M(L^0) \else $M(L^0)$\fi}}
\newcommand{\lz}{{\ifmmode L^0 \else $L^0$\fi}}
\newcommand{\ev}{{\ifmmode GeV/c^2 \else $GeV/c^2$\fi}}
\newcommand{\tri}{{\ifmmode \triangleup \else $\triangleup$\fi}}
\newcommand{\unl}{{\ifmmode U_{lL^0} \else $U_{lL^0}$\fi}}\newcommand{\gL}{{\ifmmode g_L \else $g_{L}$\fi}}
\newcommand{\gR}{{\ifmmode g_R  \else $g_{R}$\fi}}
\newcommand{\gumu}{{\ifmmode \gamma^{\mu} \else $\gamma^{\mu}$\fi}}
\newcommand{\gunu}{{\ifmmode \gamma^{\nu} \else $\gamma^{\nu}$\fi}}
\newcommand{\gdmu}{{\ifmmode \gamma_{\mu} \else $\gamma_{\mu}$\fi}}
\newcommand{\gdnu}{{\ifmmode \gamma_{\nu} \else $\gamma_{\nu}$\fi}}
\newcommand{\stw}{{\ifmmode\sin^2\theta_W \else $\sin^{2}\theta_{W}$ \fi}}
\newcommand{\sws}{{\ifmmode \;\sin^2\theta_W  \else $\;\sin^{2}\theta_{W}$ \fi}}
\newcommand{\cws}{{\ifmmode \;\cos^2\theta_W  \else $\;\cos^{2}\theta_{W}$ \fi}}
\newcommand{\sw}{{\ifmmode \;\sin\theta_W  \else $\sin\theta_{W}$ \fi}}
\newcommand{\cw}{{\ifmmode \;\cos\theta_W  \else $\;\cos\theta_{W}$ \fi}}
\newcommand{\tw}{{\ifmmode \;\tan\theta_W  \else $\;\tan\theta_{W}$ \fi}}
\newcommand{\qq}{{\ifmmode q\overline{q} \else $q\overline{q}$\fi}}
\newcommand{\lR}{{\ifmmode l_R \else $l_R$\fi}}
\newcommand{\lL}{{\ifmmode l_L \else $l_L$\fi}}
\newcommand{\nt}{{\ifmmode \nu_{\tau} \else $\nu_{\tau}$\fi}}
\newcommand{\nuR}{{\ifmmode \nu_R  \else $\nu_R$\fi}}
\newcommand{\nuL}{{\ifmmode \nu_L  \else $\nu_L$\fi}}
\newcommand{\qR}{{\ifmmode g_R  \else $q_R$\fi}}
\newcommand{\qL}{{\ifmmode q_L  \else $q_L$\fi}}
\newcommand{\qRp}{{\ifmmode q_R'  \else $q_{R}$'\fi}}
\newcommand{\qLp}{{\ifmmode q_L'  \else $q_{L}$'\fi}}
\newcommand{\est}{{\ifmmode e^{\bf \ast} \else $e^{\bf \ast}$\fi}}
\newcommand{\lst}{{\ifmmode l^{\bf \ast} \else $l^{\bf \ast}$\fi}}
\newcommand{\must}{{\ifmmode \mu^{\bf \ast} \else $\mu^{\bf \ast}$\fi}}
\newcommand{\taust}{{\ifmmode \tau^{\bf \ast} \else $\tau^{\bf \ast}$ \fi}}
\newcommand{\pperp}{{\ifmmode p_t  \else $p_t$\fi}}
\newcommand{\et}{{\ifmmode E_t  \else $E_t$\fi}}
\newcommand{\xt}{{\ifmmode x_t  \else $x_t$\fi}}
\newcommand{\smumu}{{\ifmmode \sigma_{\mu\mu}  \else $\sigma_{\mu\mu}$ \fi}}
\newcommand{\eg}{{\ifmmode e\gamma  \else $e\gamma$\fi}}
\newcommand{\epem}{{\ifmmode e^+e^-  \else $e^+e^-$\fi}}
\newcommand{\lplm}{{\ifmmode L^+L^-  \else $L^+L^-$\fi}}
\newcommand{\pp}{{\ifmmode p\overline p  \else $p\overline p$\fi}}
\newcommand{\llz}{{\ifmmode L^0\overline{L}^0 \else $L^0\overline{L}^0$\fi}}
\newcommand{\epemt}{{\ifmmode e^+e^- \to  \else $e^+e^- \to$\fi}}
\newcommand{\eb}{{\ifmmode E_{beam}  \else $E_{beam}$\fi}}
\newcommand{\ip}{{\ifmmode pb^{-1}  \else $pb^{-1}$\fi}}
\newcommand{\upm}{{\ifmmode ^{\pm}  \else $^{\pm}$\fi}}
\newcommand{\de}{{\ifmmode ^{\circ}  \else $^{\circ}$ \fi}}
\newcommand{\appr}{{\ifmmode \sim \else $\sim$ \fi}}
\newcommand{\corresp}{{\ifmmode \stackrel{\wedge}{=} \else $\stackrel{\wedge}{=}$ \fi}}
\newcommand{\sqrts}{{\ifmmode \sqrt{s} \else $\sqrt{s}$\fi}}
\newcommand{\zz}{{\ifmmode Z^0  \else $Z^0$\fi}}
\newcommand{\mz}{{\ifmmode M_{Z}  \else $M_{Z}$\fi}}
\newcommand{\mzs}{{\ifmmode M_{Z}^2  \else $M_{Z}^2$\fi}}
\newcommand{\mw}{{\ifmmode M_{W}  \else $M_{W}$\fi}}
\newcommand{\mws}{{\ifmmode M_{W}^2  \else $M_{W}^2$\fi}}
\newcommand{\mh}{{\ifmmode M_{Higgs}  \else $M_{Higgs}$\fi}}
\newcommand{\gt}{{\ifmmode \Gamma_{tot} \else $\Gamma_{tot}$\fi}}
\newcommand{\msusy}{{\ifmmode M_{SUSY}  \else $M_{SUSY}$\fi}}
\newcommand{\msusys}{{\ifmmode M_{SUSY}^2  \else $M_{SUSY}^2$\fi}}
\newcommand{\su}{{\ifmmode SU(3)_C\otimes\- SU(2)_L\otimes\- U(1)_Y \else $SU(3)_C\otimes SU(2)_L\otimes U(1)_Y$\fi}}
\newcommand{\suthree}{{\ifmmode SU(3)_C  \else $SU(3)_C$\fi}}
\newcommand{\sutwo}{{\ifmmode  SU(2)_L\otimes U(1)_Y \else $SU(2)_L\otimes U(1)_Y$\fi}}
\newcommand{\taup} {{\ifmmode \tau_{proton} \else $\tau_{proton}$\fi}}
\newcommand{\as}{{\ifmmode \alpha_{s}  \else $\alpha_{s}$\fi}}
\newcommand{\mgut}{{\ifmmode M_{GUT}  \else $M_{GUT}$\fi}}
\newcommand{\mguts}{{\ifmmode M_{GUT}^2  \else $M_{GUT}^2$\fi}}
\newcommand{\mze} {{\ifmmode m_0        \else $m_0$\fi}}
\newcommand{\mha}{{\ifmmode m_{1/2}    \else $m_{1/2}$\fi}}
\newcommand{\mb} {{\ifmmode m_{b}    \else $m_{b}$\fi}}
\newcommand{\mt} {{\ifmmode m_{t}    \else $m_{t}$\fi}}
\newcommand{\mts} {{\ifmmode m_{t}^2    \else $m_{t}^2$\fi}}

\newcommand{\mtau}{{\ifmmode m_{\tau}  \else $m_{\tau}$\fi}}
\newcommand{\dpp}{{\ifmmode \delta_{pert} \else $\delta_{pert}$\fi}}
\newcommand{\dnp}{{\ifmmode\delta_{non-pert}\else$\delta_{non-pert}$\fi}}
\newcommand{\dew}{{\ifmmode \delta_{\rm EW}\else $\delta_{\rm EW}$\fi}}
\newcommand{\rt}{{\ifmmode R_{\tau}  \else $R_{\tau} $\fi}}
\newcommand{\rz}{{\ifmmode R_{Z}  \else $R_{Z} $\fi}}

\newcommand{\swb}{{\ifmmode \sin^2\theta_{\overline{MS}} \else $\sin^2\theta_{\overline{MS}}$\fi}}
\newcommand{\cwb}{{\ifmmode \cos^2\theta_{\overline{MS}} \else $\cos^2\theta_{\overline{MS}}$\fi}}

\begin{document}

\begin{center}
\Large \textbf{Constraints on Supersymmetry from Relic Density compared with future Higgs Searches at the LHC}

\vspace{10mm}

\large

C. Beskidt$^1$, W. de Boer$^{1}$, T. Hanisch$^1$, E. Ziebarth$^1$, V. Zhukov$^{1}$, D.I. Kazakov$^{2,3}$

\normalsize
\vspace{5mm}
$^1$ \textit{Institut f\"ur Experimentelle Kernphysik,
Karlsruhe Institute of Technology,\\ P.O. Box 6980, 76128 Karlsruhe, Germany}

\vspace{5mm}
$^2$ \textit{Bogoliubov Laboratory of Theoretical Physics, Joint Institute for Nuclear Research,\\
141980, 6 Joliot-Curie, Dubna, Moscow Region, Russia}

\vspace{5mm}
$^3$ \textit{Institute for Theoretical and Experimental Physics,\\
117218, 25 B.Cheremushkinskaya, Moscow, Russia}

\vspace{30mm} \textbf{Abstract} \vspace{5mm}

\begin{minipage}[c]{12cm}

\textit{
Among the theories beyond the Standard Model (SM) of particle physics Supersymmetry (SUSY) provides
an excellent dark matter (DM) candidate, the neutralino.  One clear prediction of cosmology is
the annihilation cross section of DM particles, assuming them to be a thermal relic from the early universe.
In most of the parameter space of Supersymmetry the annihilation cross section is too small compared with
the prediction of cosmology.  However, for large values of the $\tb$ parameter the annihilation
through s-channel pseudoscalar Higgs exchange yields the correct relic density in
practically the whole range of possible SUSY masses up to the few TeV range.
The required values of $\tb$ are typically around 50, i.e. of the order of the top and bottom mass ratio,
which happens to be also the range allowing for Yukawa unification in a Grand Unified Theory
with gauge coupling unification.\\
For such large values of $\tb$ the associated production of the heavier Higgses
 is enhanced by three orders of magnitude
and might be observable as one of the first hints of new physics at the LHC.
}

\end{minipage}
\end{center}

\thispagestyle{empty}
\setcounter{page}{0}

\section{Introduction}

Cold Dark Matter (CDM) makes up 23\% of the energy of the universe,
as deduced from the temperature anisotropies in the Cosmic Microwave
Background (CMB) in combination with data on the Hubble expansion
and other observations~\cite{wmap}. One of the
most popular CDM candidates is the neutralino, a stable neutral
particle predicted by Supersymmetry (SUSY)\cite{lspdm,jungman,kolb}. The lightest
neutralinos are spin 1/2 Majorana particles, which can annihilate
into pairs of Standard Model (SM) particles. They are usually the Lightest Supersymmetric
Particles (LSPs), which are stable if R-parity is conserved.  This multiplicative quantum number  forbids decays of SUSY particles with negative R-parity to known particles with positive R-parity, while the Next-to-Lightest SUSY Particles (NLSPs) and heavier ones will decay to the LSP and SM particles.

SUSY is the most popular candidate for a theory beyond the SM, since it is the only theory so far, which solves several problems simultaneously: it provides a splendid
candidate for DM, it has no quadratic divergencies in the Higgs sector, it allows for unification of gauge and Yukawa couplings and it predicts the Higgs mechanism by radiative corrections\cite{susyrev}.
Search for SUSY particles is one of the prime objectives of the new Large Hadron Collider (LHC), which just
starting taking data with several detectors\cite{lhc}.
But even if SUSY would be discovered at the LHC, it does not guarantee that the DM in the universe would be made of neutralinos, since many other candidates exist, see e.g. \cite{Bertone:2004pz}.

To prove that the DM is really made of neutralinos one would need to correlate additional properties,
like e.g. the mass or the annihilation cross section. The latter is a particularly good property, if
DM is thermal relic from the early universe, since it is directly correlated with the relic density in an
almost model independent way from the expansion history of the universe\cite{jungman,kolb}. Several studies exist on how well one can determine the annihilation cross section from accelerator data, see e.g. \cite{Nath:2010zj} and references therein. Up to now only methods in the so-called bulk region, focus point region or co-annihilation regions have been considered, since here rather specific SUSY mass spectra exist. In the so-called funnel region, where the annihilation proceeds through s-channel exchange of the pseudoscalar Higgs $A$, this has not been studied, since it is usually considered as a narrow stripe in parameter space for which the neutralino mass is close to the A-resonance, i.e. $m_\chi \approx m_A/2$.
Of course, one can consider $m_A$ to be a free parameter and tune the mass to get the correct relic density
or for that matter tune the masses of other SUSY particles contributing to annihilation. In this case the relic density does not constrain. However, in specific SUSY breaking scenarios all SUSY masses are related to each other
via renormalization group equations\cite{susyrev}. In this Letter we consider the popular mSUGRA model with unified breaking scales for the fermions and bosons, respectively. This scenario allows for unification of gauge and Yukawa couplings and radiative electroweak symmetry breaking by minimizing the Higgs potential. In this case the mass of the pseudoscalar Higgs is sensitive to $\tb$, the ratio of vacuum expectation values of the neutral components of the two Higgs doublets. Since $\tb$ is a free parameter one expects that the correct relic density or equivalently the annihilation cross section can be obtained in almost any region of parameter space by tuning $\tb$. It is the purpose of the present letter to see if this is true and discuss this in the context of future LHC physics.

As it turns out, in a large region of parameter space the value of $\tb$ is required to be around 50, which implies an enhancement of the
pseudoscalar Higgs production at the LHC  by three orders of magnitude, since this cross section is $\propto \tan^2\beta$\cite{djouadi}.
 It should be noted that for these large values of $\tb$ the contribution from other annihilation channels can be neglected, so discovering the Higgs at the LHC allows for the possibility to determine the relic density from LHC data and thus establishing a connection between the DM in the universe and the neutralinos produced in laboratory experiments.
The value of 50 is furthermore of interest, since it is of the
order of the ratio of top- and bottom-quark masses, which occurs naturally in SO(10) GUT theories. They feature in addition gauge coupling unification\cite{amaldi,bs}, Yukawa coupling unification\cite{deBoer:2001nu,so10}  and right-handed neutrinos, which are needed to generate small neutrino masses via the see-saw mechanism\cite{seesaw}.

 Several heavy flavour physics observables  are enhanced by large values of $\tb$  as well (see e.g. \cite{Isidori:2007jw} and references therein).
Especially, the FCNC decay of $B_s\rightarrow \mu\mu$, which proceeds in the SM via loops involving top quarks and W-bosons, is enhanced in SUSY by $\tan^6\beta$. Experimental constraints from this $\tb$ sensitive variable will be discussed, but excluded regions in SUSY parameter space by other low energy observables
are outside the scope of this paper, since  deviations from the SM are at   the 2-3$\sigma$ level.   The errors are dominated by statistical and systematic uncertainties, so the  excluded or preferred regions
depend on the assumed errors and the applied statistical treatment, which results in  large differences between the different analysis, see e.g. for some recent analyses Refs. \cite{Buchmueller:2009fn,Trotta:2008bp,Akrami:2009hp,Feroz:2008wr} and references therein. As said, discussing these analyses is beyond the scope of this letter, which concentrates on the SUSY parameter space allowing for a  DM relic density determination from  Higgs production at the LHC. The precision with which this can be done depends on the LHC luminosity and will be studied in a future paper.

\begin{figure}[]
\begin{center}
\includegraphics[width=0.7\textwidth]{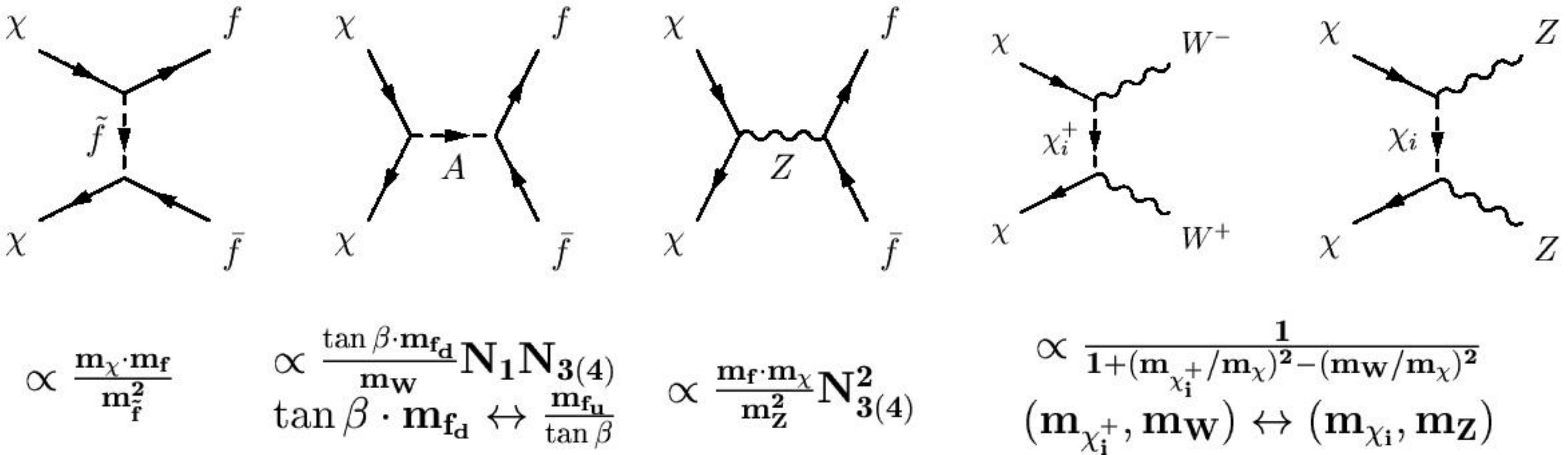}
\caption[]{ Annihilation diagrams for  the lightest neutralino, which is
  a linear combination of the gaugino and Higgsino states:
  $|\chi_o>=N_1 |B_0>+N_2|W^3_0>+N_3|H_1>+N_4|H_2>$. The dependence of the amplitudes on  masses
  and neutralino mixing parameter $N_i$ has been indicated. }
\label{f1}
\end{center}
\end{figure}
\section{Predictions from mSUGRA}
\label{susywimp}
The relic density and annihilation cross section $\sigma$ are related through:
\begin{equation}
\Omega h^2=\frac{3.10^{-27}}{<\sigma v>},
\label{e1}
\end{equation}
where the annihilation cross section $\sigma$ averaged over the relative velocities of the neutralinos is given in pb \cite{jungman} and $h\approx 0.71$ is the Hubble constant in units of 100 $(km/s)/Mpc$\cite{wmap}.  The best value for the relic density is $\Omega h^2=0.1131\pm 0.0034$ \cite{wmap}.  For the relation between cross section and expansion rate in Eq. \ref{e1} one assumes DM was a thermal relic, which froze out at the time when the annihilation rate was about equal to the expansion rate, given by the Hubble constant.
For a given relic density $\Omega$ the annihilation cross section is known
independent of a specific model, since it only depends on the observed Hubble constant and the observed relic density. Its value is furthermore largely independent of the neutralino mass $m_\chi$ (except for logarithmic corrections)\cite{jungman,kolb}.
The DM constraint should exist for any model, but to be specific  the
 mSUGRA model, i.e. the Minimal Supersymmetric Standard Model (MSSM) with supergravity
inspired breaking terms, will be considered \cite{susyrev}. It is characterized by only 5 parameters:
$m_0,~m_{1/2},~\tb,~\mbox{sign}(\mu), ~A_0$. Here $m_0$ and $m_{1/2}$ are
the common masses for the gauginos and scalars at the GUT scale, which is determined by
the unification of the gauge couplings at this scale. Gauge unification is perfectly possible with the
latest measured couplings at LEP~\cite{bs}.
We only consider the dominant trilinear
couplings of the third generation of quarks and leptons and set its common value at the GUT scale
 $A_0$ equal zero, but the low energy values are different for all generations because of the
 different radiative corrections.
 $A_0$ values different from zero were found not to influence the relic density, but are
 important if the relic density is considered in combination with $B_s\rightarrow \mu\mu$.
Electroweak symmetry breaking (EWSB) fixes the scale of $\mu$
\cite{susyrev}, so only its sign is a free parameter.
  The positive sign is taken, as
suggested by the small deviation of the SM prediction from the muon anomalous moment\cite{deBoer:2001nu}.

The Born-level neutralino annihilation diagrams
are shown in Fig. \ref{f1}. The cross sections  are
proportional to the final state fermion mass, which originates either from the Yukawa
couplings for the Higgs exchange diagram or from the helicity suppression at the low
energies involved in cold DMA \cite{goldberg}. Therefore heavy fermion final states, i.e.
third generation quarks and leptons, are expected to be dominant. The Higgs exchange
diagram is in addition proportional to \tb ~ for down type quarks and 1/\tb ~ for up type
quarks, so top quark final states
are suppressed for large \tb~ and the main branching ratios will be into b-quarks, tau-leptons and muons.
The W- and Z-final states have  usually a much smaller cross
section due to the weak couplings involved, as was explicitly calculated with the CalcHEP program\cite{calchep}.
\begin{figure}[]
\begin{center}
\includegraphics[width=0.4\textwidth]{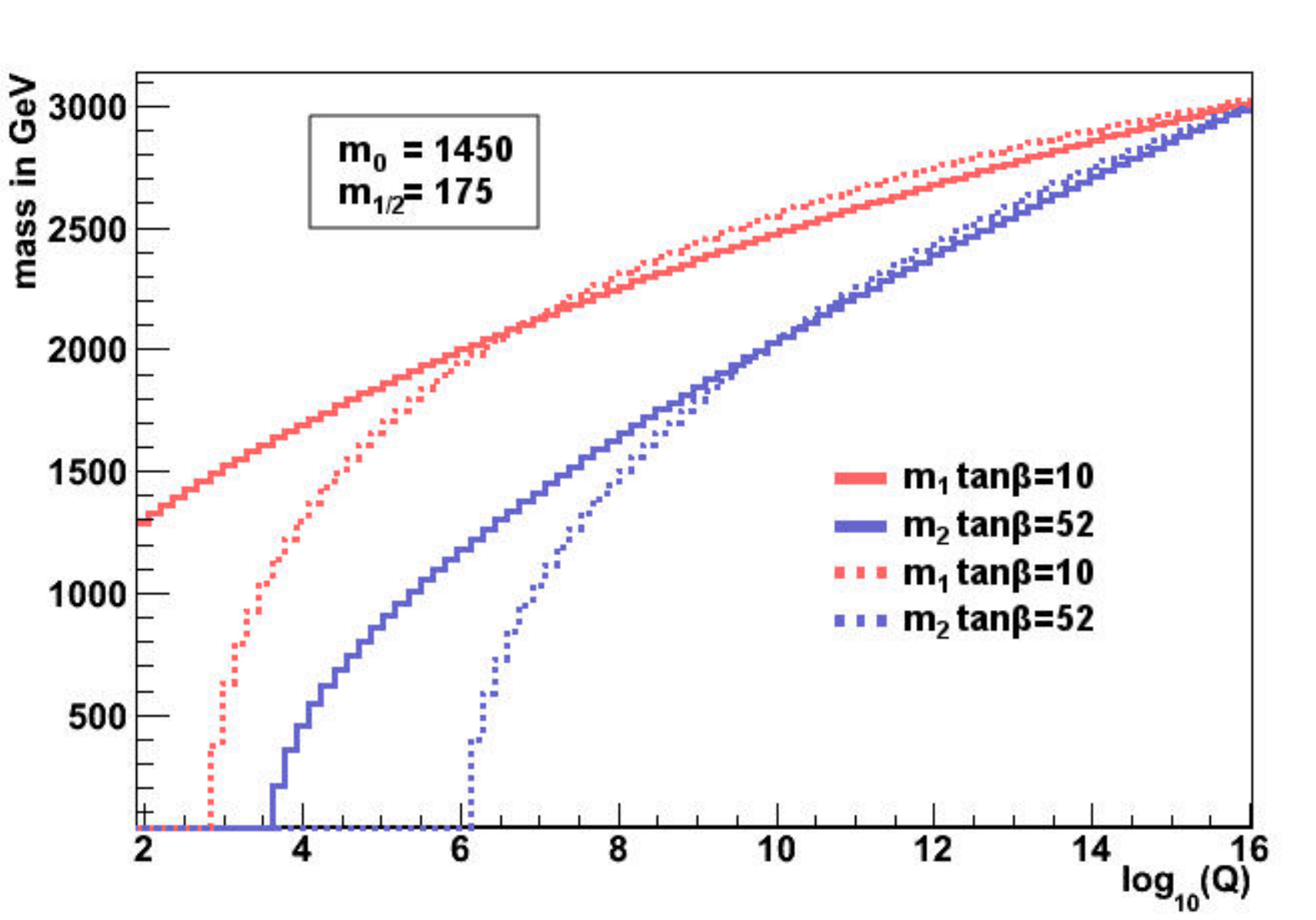}
\includegraphics[width=0.4\textwidth]{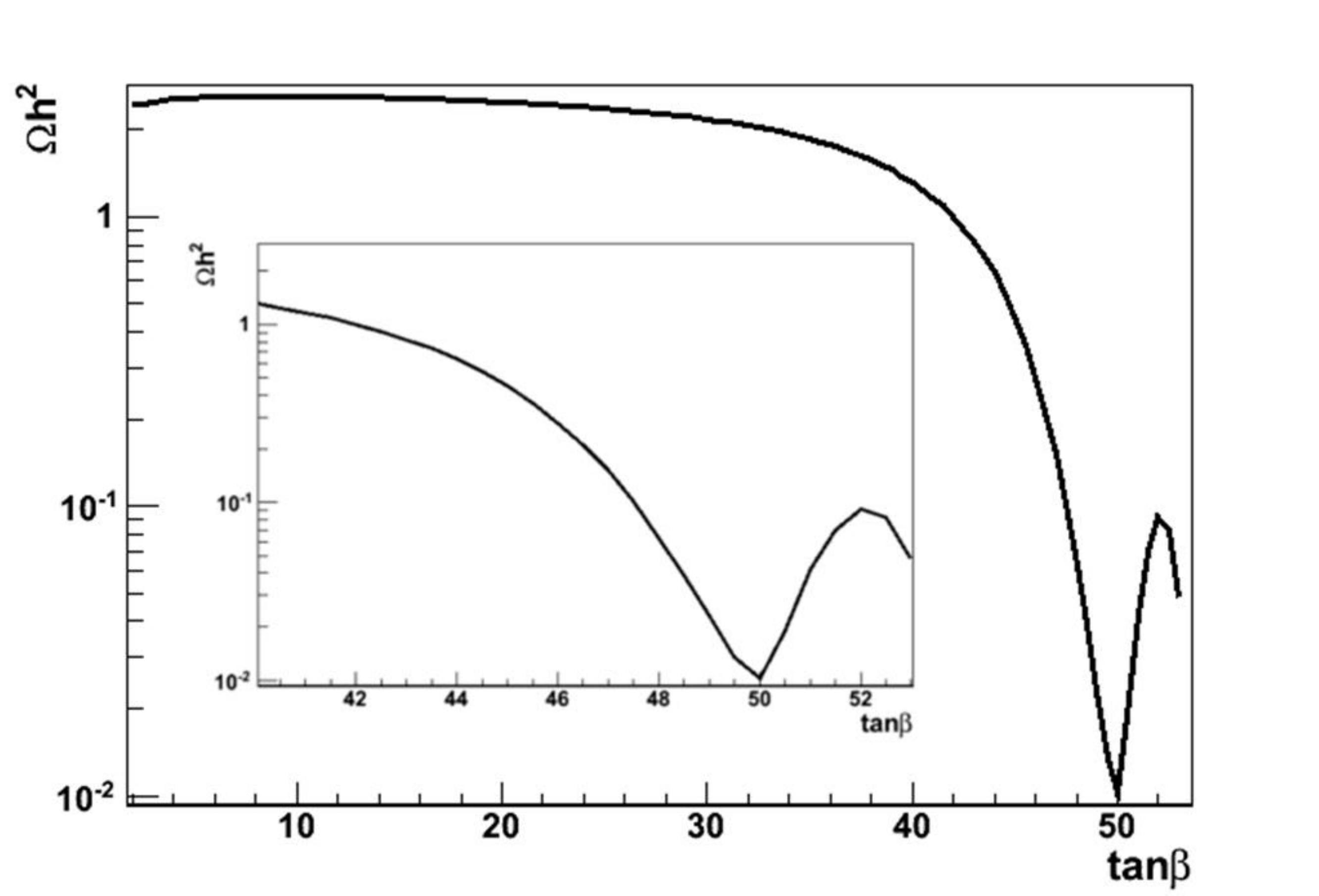}
\caption{Left:Running of Higgs mass terms for different values of $\tan\beta$ for $m_0=1450$ and $m_{1/2}=175$ GeV. One observes that for
large $\tan\beta$ the pseudoscalar mass, which at tree level equals $m_1^2+m_2^2$,
becomes small for large $\tan\beta$,
so for large $\tan\beta$ the dominant annihilation channel is via pseudoscalar Higgs exchange.
Right: The strong dependence of the relic density as
function of $\tan\beta$. }
\label{f2}
\end{center}
\end{figure}
In the MSSM the two mass terms of the two Higgs doublets, denoted by $m_2$ and $m_1$, receive radiative corrections from the top- and bottom Yukawa couplings, respectively \cite{susyrev}. The common value of these masses at the unification scale is $\sqrt{m_0^2+\mu_0^2}$, but the large top Yukawa coupling can drive $m_2^2$ negative at the electroweak scale for heavy top masses, thus leading to radiative electroweak symmetry breaking (EWSB). The top mass was predicted in this way to be heavy well before the top discovery\cite{ewsb}. For  values of $\tb\approx m_t(m_t)/m_b(m_t)\approx 160/3\approx 50$ the bottom Yukawa coupling becomes of the same order as the top Yukawa coupling, which could lead to unification of all Yukawa couplings of the third generation\cite{so10}. But in this case also $m_1$ is driven close to zero as well or becomes negative, as
 demonstrated on the left-hand side of  Fig. \ref{f2}. 
 This implies that the pseudoscalar Higgs mass, which at tree level is given by $m_A^2=m_1^2+m_{2}^2$, becomes light for large $\tb$.
Given that the relic density by $A$ exchange is proportional to $1/(4m_\chi^2-m_A^2)^2$ and $m_A$ decreases fast with increasing $\tb$ the relic density is a strong function of $  \tb$, as demonstrated on the
right-hand side of  Fig. \ref{f2}. The value of $\tb$ to obtain the WMAP relic density was calculated for each point in the $m_0-m_{1/2}$ plane by simply scanning $\tb$. The relic density was calculated with the public code micrOMEGAs 2.4\cite{micromegas} using Suspect 2.41 as mSUGRA mass spectrum calculator\cite{suspect}.

\begin{figure}[]
\begin{center}
\includegraphics[width=0.4\textwidth]{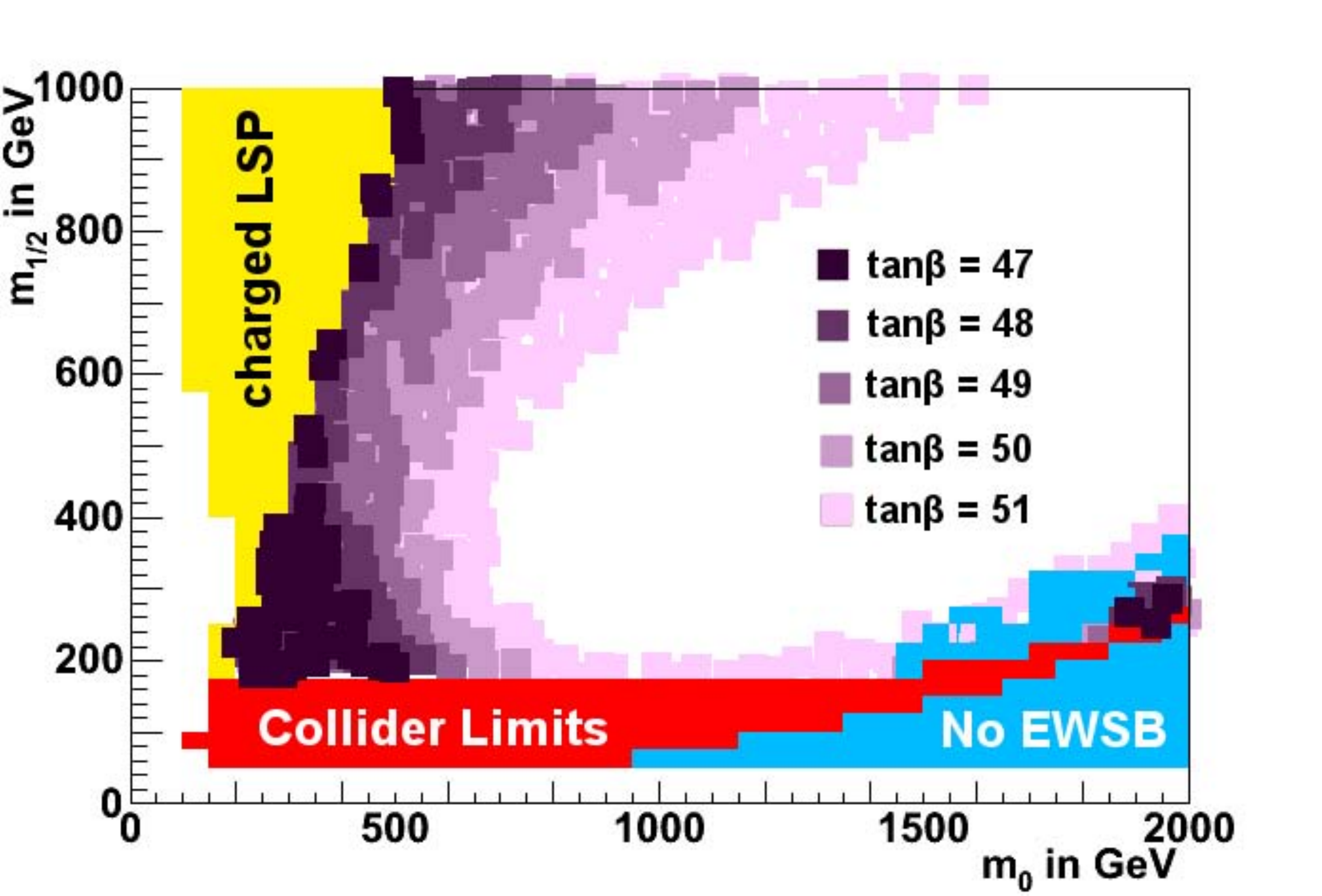}
\includegraphics[width=0.4\textwidth]{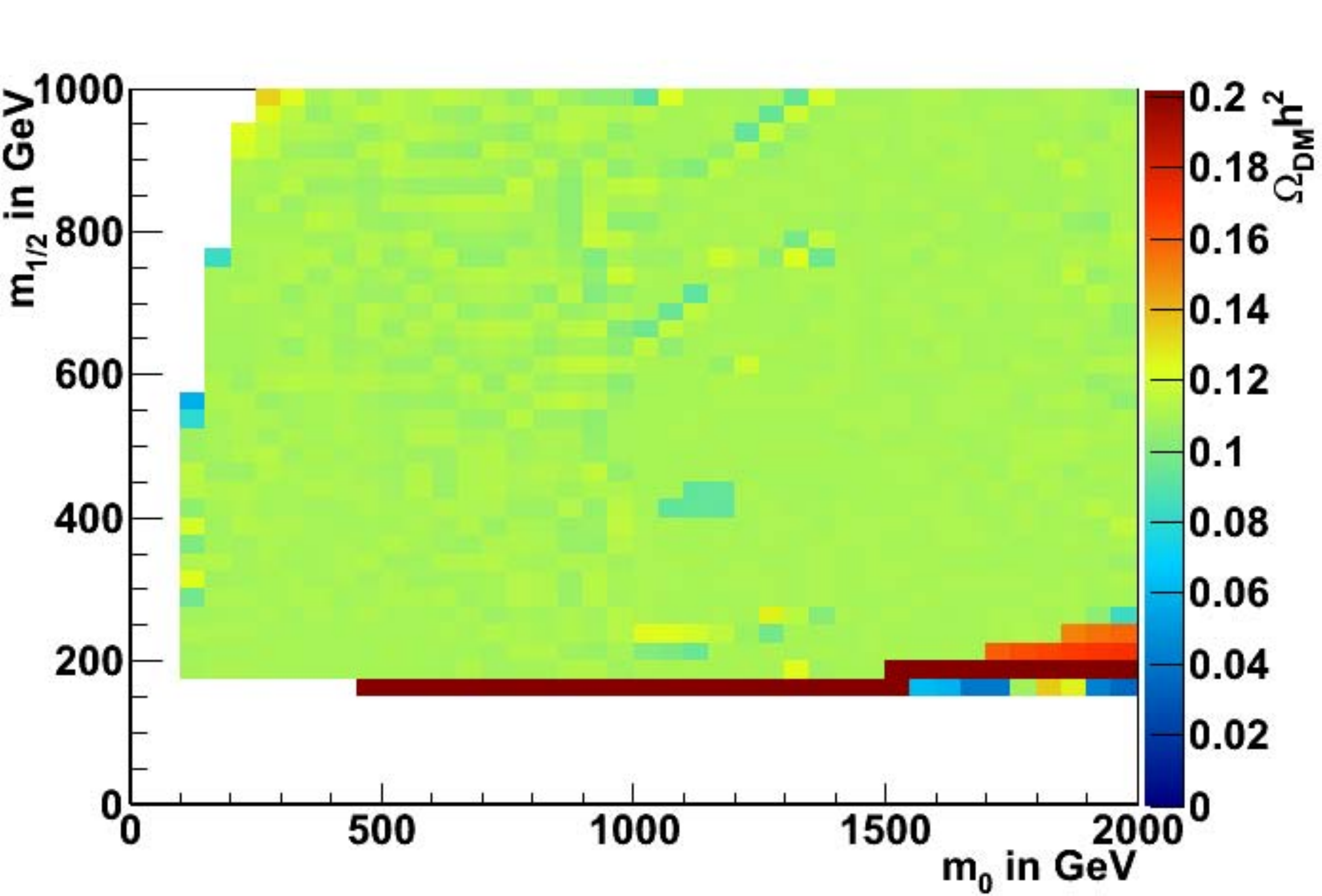}
\caption[]{
Left: The region for correct relic density for various values of $\tan\beta$. One observes that the "funnel" region varies  with $\tb$. The trilinear couplings are 0 at the GUT scale and the sign of $ \mu$=+1.
Right:  Relic density in the $m_0-m_{1/2}$ plane after tuning the value of $\tan\beta$. 
}
\label{f3}
\end{center}
\end{figure}

\begin{figure}[]
\begin{center}
\includegraphics[width=0.4\textwidth]{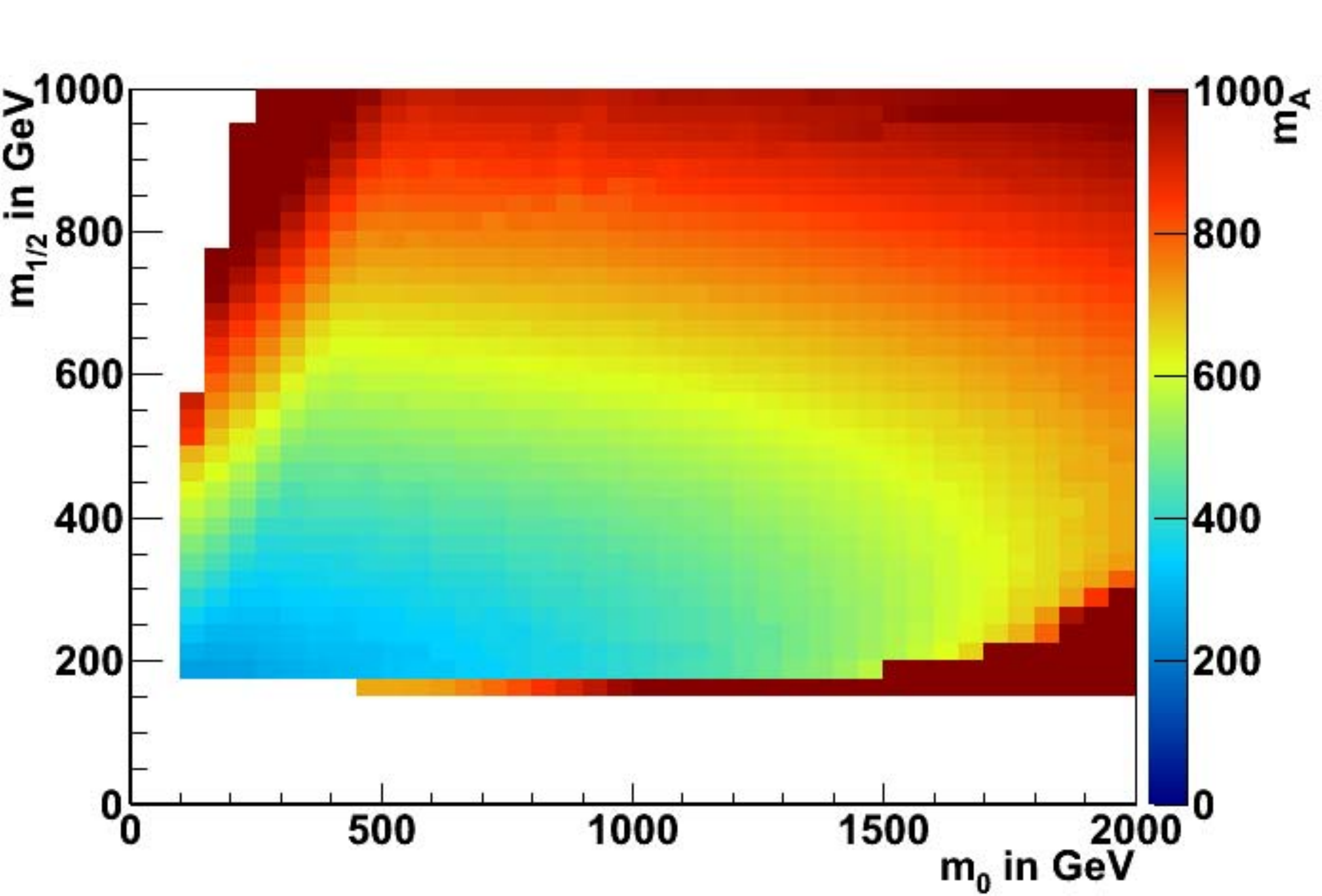}
\includegraphics[width=0.4\textwidth]{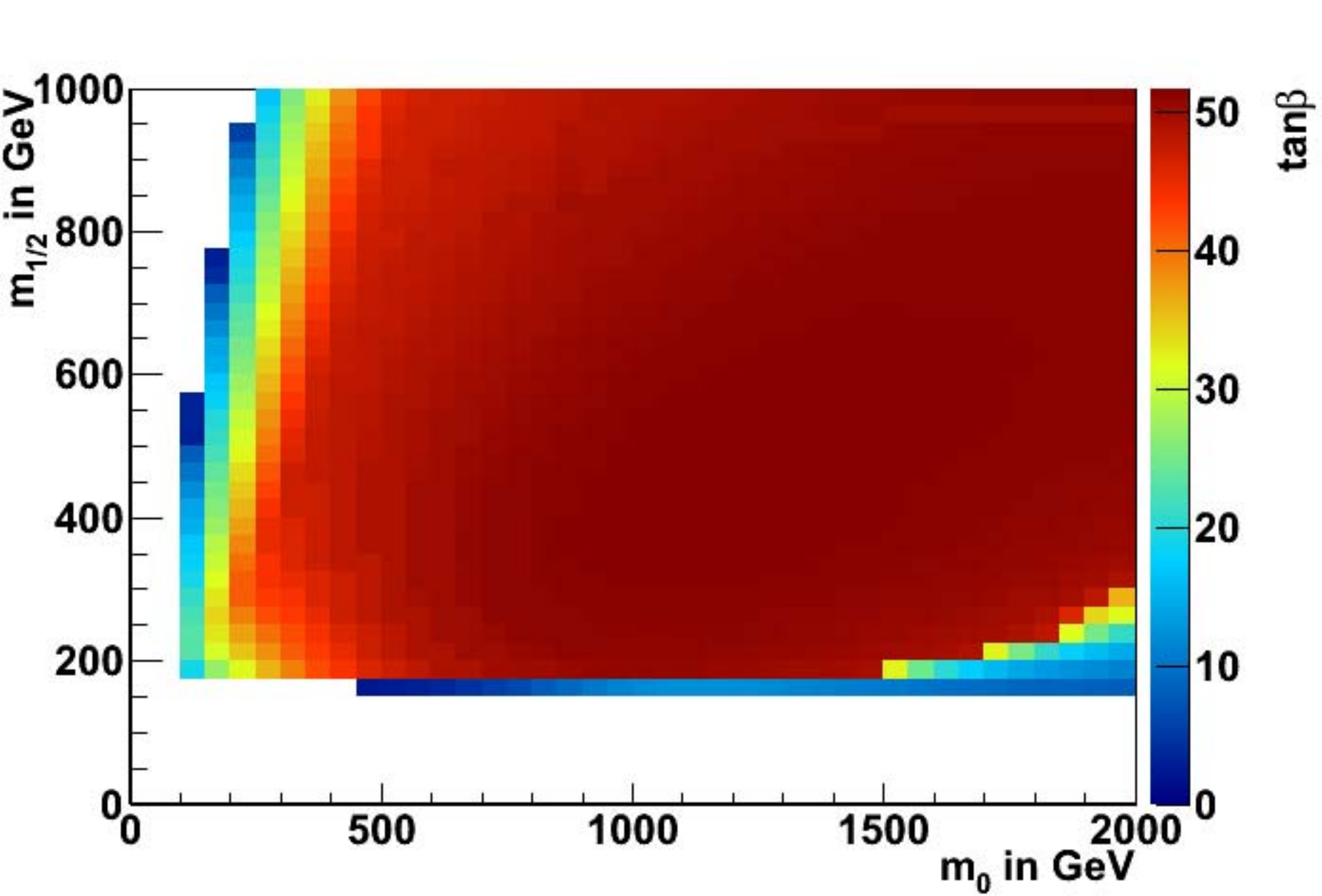}
\caption{
The values of $m_A$ (left) and $   \tan   \beta$ needed for the correct relic density in Fig.  \ref{f3} (right).
}
\label{f4}
\end{center}
\end{figure}

Figs. \ref{f3} and \ref{f4} shows that $\tb$ can be adjusted in the whole $m_0-m_{1/2}$ plane to get a correct relic density: the right-handed panels show the relic density and the values of $\tb$ needed for the correct relic density.
  As mentioned in the introduction low energy SUSY masses are excluded, since else the effect of SUSY in loops would have shown up at LEP or heavy flavour observables.
  Since $Br(B_s\rightarrow \mu\mu) $ is proportional to $\tan^6\beta$ it is an important constraint for large $\tb$ scenarios. But this observable is also sensitive to the trilinear coupling at the GUT scale $A_0$ and
  if $A_0$ is left free,  both the relic density constraint and the 90\% C.L. upper limit of $Br(B_s\rightarrow \mu\mu) < 4.7\cdot 10^{-8}$ \cite{pdg} can be met in the whole plane and
  no excluded region from $Br(B_s\rightarrow \mu\mu)$ is observed.

The  values of $m_A$   for the correct relic density  are shown on the left-hand panel of Fig. \ref{f4}. One observes that for larger neutralino masses, i.e. larger values of $m_{1/2}$, the values of $m_A$ increase, as expected, since the neutralinos have to be on the tail of the pseudoscalar Higgs resonance for a correct relic density, typically $m_A/2m_\chi\approx 1.2-2.4 $.
It is interesting to note, that the needed values of $\tb$ are in practically the whole region close to 50, as expected since in that case the Yukawa coupling
of the bottom quark becomes of the same order as the one of the top quark, so both Higgs mass terms  in Fig. \ref{f2} become small.

In Fig. \ref{f3} the top left is excluded, since here the LSP is not a neutral particle, but the stau is the LSP.
The bottom right is excluded, since here EWSB does not work, since  $\mu^2$ becomes negative, where $\mu$ is the mixing parameter in the Higgs potential. In the transition to these forbidden regions
co-annihilation of the LSP with an almost mass-degenerate other SUSY particle is possible. The co-annihilation and other properties of these transition regions have been reviewed nicely in Ref. \cite{nanopoulos} and the reader is referred to this reference for details and original references.
Here only the salient features of these regions important for this analysis are mentioned.
  In the transition to the forbidden region at the top left the staus and lightest neutralinos are nearly degenerate in mass, so they will freeze out in the early universe at the same temperature. This leads to co-annihilation of the stau and neutralino into tau leptons, thus  reducing the relic density according to Eq. \ref{e1}.
In the bottom right-hand corner a similar feature is observed: here EWSB does not work, while $\mu^2$ becomes negative and the lightest chargino and neutralino become almost degenerate. This can be understood as follows: the starting point of the Higgs mass terms  at the GUT scale  (see  Fig. \ref{f2}) is $\sqrt{m_0^2+\mu^2}$, so for large values of $m_0$, $\mu$ has to become small. Else the Higgs mass terms in Fig. \ref{f2} will not become negative before the electroweak scale, as required for EWSB. If $\mu$ is small, the LSP and chargino mass terms are both determined by the smallest diagonal matrix element in the mass matrices, which is $ \mu$ in this case. So in the transition region chargino-neutralino become mass-degenerate and their co-annihilation into W-bosons dominates. Then $\tb$ has to be reduced again in order not to have a too large annihilation cross section, as is clear from the low $\tb$ regions in the right-hand panel of Fig. \ref{f4}.

Finally, at small values of $m_0$ and $m_{1/2}$ the sfermions are light and the left-hand diagram of Fig. \ref{f1} dominates and again $\tb$ has to be reduced in order not to have a too large annihilation cross section.
This region is traditionally called the bulk region, since it does not correspond to a narrow stripe.

In summary, if one allows $\tb$ to vary in the $m_0-m_{1/2}$ plane, one obtains the observed relic density for $any$ combination of $m_0$ and $m_{1/2}$, i.e. the relic density allows $all$ masses for the SUSY   sparticles.
Such an mSUGRA scenario with large $\tb$ can be uniquely tested at the LHC because of the large cross section for  heavy Higgs production, as will be discussed in
the next section.

\section{Expected Constraints from Higgs Production at the LHC}
\begin{figure}[]
\begin{center}
\includegraphics[width=0.4\textwidth]{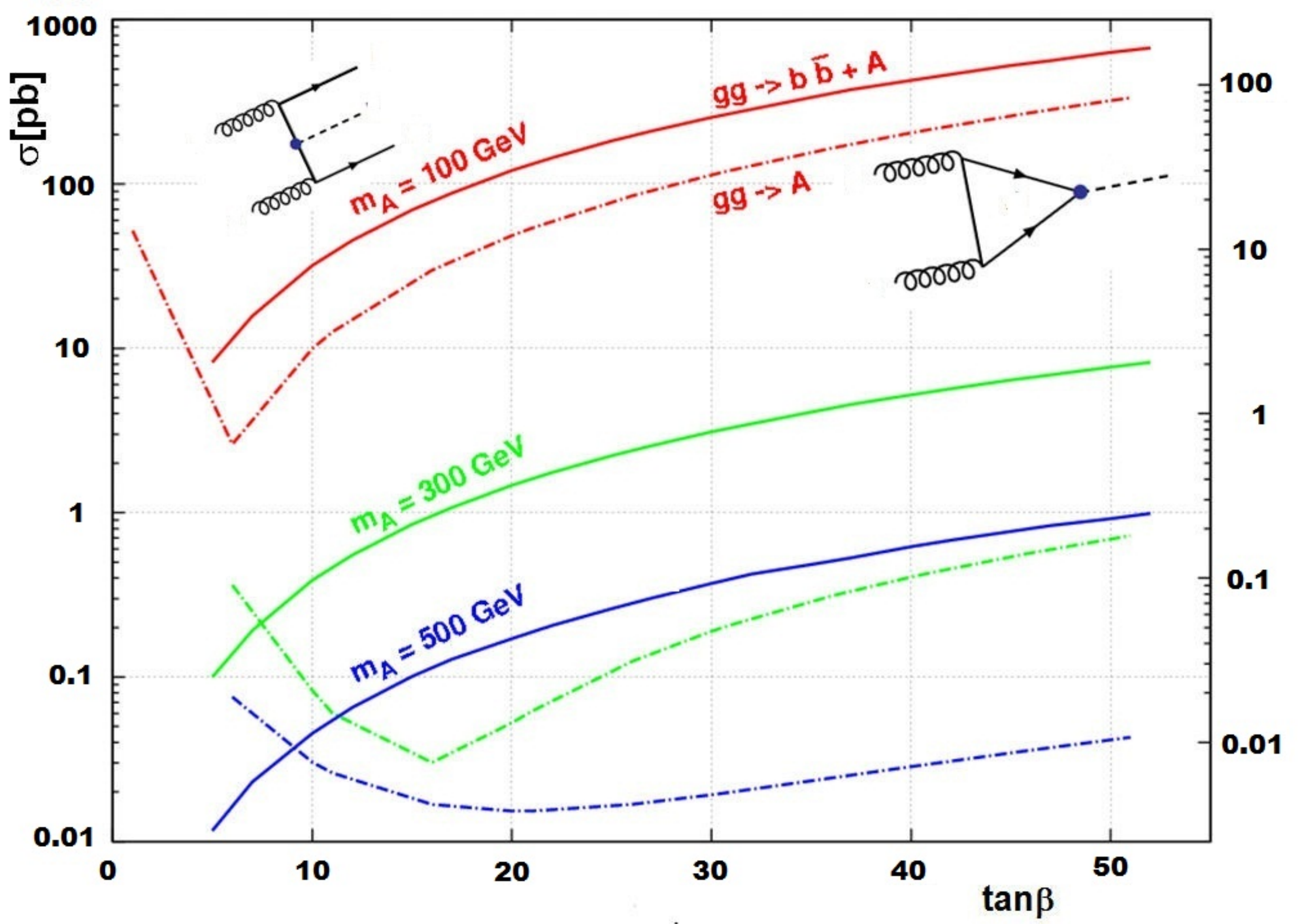}
\includegraphics[width=0.4\textwidth]{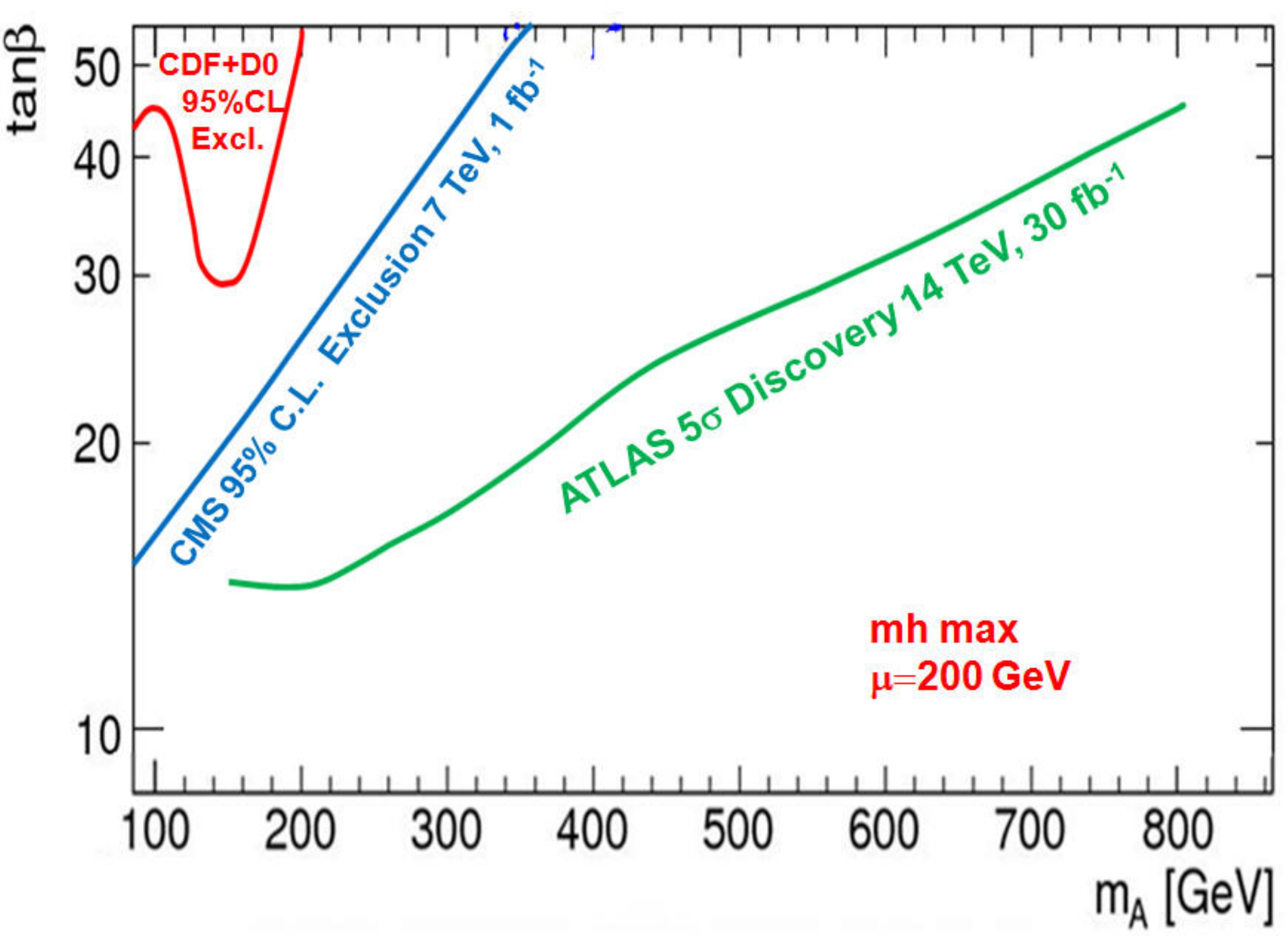}
\caption{Left:
The pseudoscalar Higgs production cross section as function of  $\tan\beta$, both for the gluon fusion diagram and associated Higgs production with a b-quark for the different Higgs mass values indicated. Since the b-quark production is mostly in the forward direction, the scale on the right-hand side indicates if at least one b-quark is required to be in the acceptance, defined by $\eta<2.5$, and have a transverse momentum above 20 GeV/c.  Right: expected discovery reach for the ATLAS detector at 14 TeV  and a luminosity of 30 $fb^{-1}$\cite{atlashiggs}. The region already excluded at the Tevatron \cite{tevatronhiggs} and the expected exclusion reach after the initial 7 TeV run  at the LHC\cite{cmshiggs} have been indicated as well (assuming a luminosity of 1 $fb^{-1}$).
These sensitivity projections  for future LHC running of the ATLAS and CMS detectors  are preliminary.
}
\label{f5}
\end{center}
\end{figure}
The MSSM has five Higgs particles: the light Higgs h with a mass below 130 GeV, two heavy neutral Higgses
(a scalar and pseudoscalar one, denoted by $H$ and $A$, respectively) and two charged Higgses\cite{higgshunter}.
If the heavy Higgses are well above the light Higgs, all masses of the other charged and neutral Higgses are nearly degenerate and the production cross sections for the heavy scalar and pseudoscalar Higgs are nearly
identical, so in practice one obtains twice the cross section, since the interference between the diagrams
with pseudoscalar and scalar Higgses  is negligible, even if the decay is into the same final state.

The cross section for Higgs production at the LHC has been extensively studied, both in the SM and MSSM.
For  references we refer to the excellent reviews by Djouadi\cite{djouadi}.
For the SM the gluon fusion via a top loop is the dominant diagram. The next dominant diagrams are the annihilation of heavy quarks into Higgs bosons. In next-to-leading order these heavy quarks, mainly bottom quarks, are obtained from gluon splitting. 
However, in the MSSM the Higgs coupling to down(up)-type quarks is proportional to the $ \tb (1/\tb)$,
so the coupling to top quarks is strongly suppressed at large $\tb$ and the production via the top loop is not  dominant anymore, but the production in association  with b-quarks dominates. This $\tb$ dependence is shown on the left-hand side of Fig. \ref{f5} for various Higgs masses. The associated production cross sections were calculated at Born level with CalcHEP\cite{calchep}, while the gluon fusion via a loop was calculated with the formulae from the Higgs Hunter's Guide\cite{higgshunter} after correcting a misprint in the squark contribution. Note that for small $\tb$ the SM behavior  with the production via the top loop dominating is recovered.

For values of $\tb=50$ the cross section for associated production of b-quarks and Higgses is enhanced by a factor 2500, so Higgs bosons can either be easily found or excluded.  Although the Higgs decay into tau final state has only a 10\% branching ratio, it is still the preferred search channel because of the strongly reduced background.
At the LHC one expects   the preliminary discovery reach for pseudoscalar Higgses to reach 800 GeV, as shown in Fig.     \ref{f5} on the right-hand side for a recent analysis by the ATLAS Collaboration \cite{atlashiggs}. Here a luminosity of 30 $fb^{-1}$  at the maximum LHC energy of 14 TeV was assumed,
which corresponds to several years of running.
Similar results were obtained earlier by the CMS Collaboration \cite{cmshiggs}. The region excluded at present by the combined data from the CDF and D0 Collaborations at the Tevatron (using 1.8-2.2 $fb^{-1}$ at a center-of-mass energy of 1.96 TeV) has been indicated as well\cite{tevatronhiggs}. The expected exclusion
for the initial LHC run at 7 TeV is indicated from the preliminary CMS study assuming a luminosity of
1$fb^{-1}$\cite{cmshiggs1}.

If one compares the right-hand side of Fig. \ref{f5}  with the $m_A$ values plotted in Fig. \ref{f4} one sees that the Higgs search is sensitive to $m_{1/2}$ scales up to 800 GeV. In mSUGRA the gluinos are 2.7$m_{1/2}$, so in this scenario with the neutralino providing all the dark matter, the Higgs search is sensitive to gluino mass scales up to 2 TeV.

\section{Conclusion}
In this Letter it has been shown that if the neutralino makes up the dark matter of the universe, the neutralino annihilation cross section, as expected from cosmology for a thermal relic, requires a value of $\tb$ of the order of 50, i.e. of the order of the ratio of top and bottom quark mass in a large region of
parameter space.

For such large values of $\tb$ the associated production of the heavier Higgses is enhanced by $\tan^2\beta$, i.e. by three orders of magnitude. Combining the expected reach of Higgs searches of about 800 GeV with the  cosmological preferred region, if one assumes the neutralino to be the dark matter candidate,  leads to a discovery reach in the mSUGRA mass plane corresponding to gluinos up to 2 TeV.
  Here we discussed the search for the pseudoscalar Higgs, but the heavy scalar Higgs has typically the same mass, thus enhancing the cross sections by a factor two, since the interference between the pseudoscalar and scalar Higgs production is negligible
(zero at Born level).

If only a fraction of the dark matter consists of neutralinos, the annihilation cross section increases
according to Eq. \ref{e1} and so for a given neutralino mass the $m_A$ mass has to be lower.
  So if $m_A$ has to be lower for a given value of $m_{1/2}$ one obtains  an
even larger reach in the $m_0-m_{1/2}$ plane by the Higgs searches.

It should be noted that we analyzed the DM constraint in the framework of mSUGRA. However, the DM annihilation depends mainly
on the masses of neutralinos, the neutralino mixing, the pseudoscalar Higgs mass and $\tb$, if one is outside the  bulk and co-annihilation regions.
So in similar scenarios with large $\tb$, as expected from Yukawa unification, the relic density constraint can be fulfilled, which implies that the Higgs production could be the first sign of new physics beyond the SM.\\[2mm]
{\bf Acknowledgements.} Support from the Deutsche Forschungsgemeinschaft (DFG) via a Mercator Professorship
(Prof. Kazakov) and the Graduiertenkolleg  "Hochenergiephysik und Teilchenastrophysik" in Karlsruhe  is
greatly appreciated. Furthermore, support from the Deutsche Luft und Raumfahrt (DLR) and the Bundesministerium
for Bildung und Forschung (BMBF) is acknowledged.

\end{document}